\documentclass[epjCONF]{svjour}
\usepackage{graphics}
\session-title{MESON 2012 - 12 International Workshop on Meson Production, Properties and Interactions}
\begin{document}
\title{Evidence of new nucleon resonances from electromagnetic meson production\footnote{Invited talk given at MESON 2012, Krakow, Poland.}}
\author{Volker D. Burkert\inst{}\thanks{\email{burkert@jlab.org}}, for the CLAS collaboration}
\institute{Jefferson Lab, 12000 Jefferson Avenue, Newport News, VA 20606, USA}
\begin{abstract}
{The study of nucleon resonances in electromagnetic meson production with the CLAS detector is discussed. The electromagnetic interaction is complementary to pion scattering in the exploration of the nucleon excitation spectrum. Higher mass states often decouple from the $N\pi$ channel and are not seen in $\pi N \to \pi N$. Photoproduction of mesons, such as $K\Lambda$, $\omega p$ and $\eta^\prime p$ may be more sensitive to many of these states. The CLAS detector, combined with the use of energy-tagged polarized photons and polarized electrons, as well as polarized targets and the measurement of recoil polarization, are the tools needed for a comprehensive nucleon resonance program. Some of the recently published high statistics data sets had significant impact on further clarifying the nucleon excitation spectrum. }
\end{abstract}
\maketitle
\section{Introduction}
\label{intro}
Understanding the systematics of the nucleon excitation spectrum is key to understanding the effective 
degrees of freedom underlying baryonic matter. The challenge is that the 
spectrum consists of broad and overlapping states that cannot easily be isolated from each other, and
excitation levels can only be predicted using models.  The most comprehensive predictions 
of the resonance excitation spectrum have come from the various implementations 
of the constituent quark model based on (broken) $SU(6) \times O(3)$ symmetry\cite{capstick}. 
Yet this model does not include gluonic degrees of freedom (hybrid baryons), or
resonances that are generated dynamically through baryon-meson interactions. 
Recent advances in Lattice QCD led to predictions of the nucleon spectrum in QCD with 
dynamical quarks\cite{dgr}, albeit with still large pion 
 masses of 396 MeV. Lattice prediction can therefore only be taken as indicative of the quantum numbers 
 of excited states and not of the masses of specific states. In parallel, the development of dynamical coupled 
 channel models is being pursued with new vigor. The EBAC group at JLab has
 shown\cite{kamano} that coupled channel effects can results in significant mass shifts of the excited states. 
 As a particularly striking result, a very large shift was found of the Roper resonance pole mass  to 1365 MeV
 downward from its bare core mass of 1736 MeV. This result has clarified the 
 longstanding puzzle of the incorrect mass ordering of $P_{11}(1440)$ and $S_{11}(1535)$ resonances in 
 the constituent quark model. The developments on the phenomenological side go hand in hand with a 
 world-wide experimental effort to produce high precision data in many different channel as a basis for 
 a largely model-independent determination of the light-quark baryon resonance spectrum for masses above 1800 MeV/c$^2$. 
 On the example of the CLAS results, the strong impact of the precise meson photoproduction data is discussed.  Several reviews have recently been published on this and related subjects\cite{klempt,aznauryan,tiator}.

\section{The search for new excited states of the nucleon}
\label{sec:1}

The complex structure of the excitation spectrum complicates the experimental search for individual states. 
Most of the excited nucleon states listed in the Review of Particle Properties prior to 2012 have 
been observed in elastic pion scattering $\pi N \to \pi N$. However there are important limitations in the 
sensitivity to the higher mass nucleon states that may have 
very small decay widths $\Gamma_{\pi N}$.  The extraction of resonance contributions for such states
then becomes exceedingly difficult. Estimates for alternative decay channels have been made in quark model 
calculations\cite{capstick} for various channels.  This has led to a major experimental effort at JLab, ELSA, GRAAL, and MAMI
to chart differential cross sections and polarization observables for a variety of meson photoproduction channels with high 
precision. At CLAS, several final states have been measured with high precision\cite{mdugger1,mdugger2,mwilliams1,mwilliams2,mwilliams3,bradford1,mcnabb,mccracken,bdey2010,bradford2}.  Here I focus on 
measurements with $K^+\Lambda$ or $K^+\Sigma^\circ$ in the final state. 
The ultimate goal of these measurements is to arrive at so-called  "complete experiments". In  $\gamma p \to K^+\Lambda$ 
there are four complex parity-conserving amplitudes 
which can be determined with 8 well-chosen measurements that include one cross section measurement and a 
combination of measurements with polarized beam, polarized target  and recoil polarization. 
\begin{figure}[t]
\begin{center}
\hspace{-0.8cm}
\resizebox{0.7\columnwidth}{!}{\includegraphics{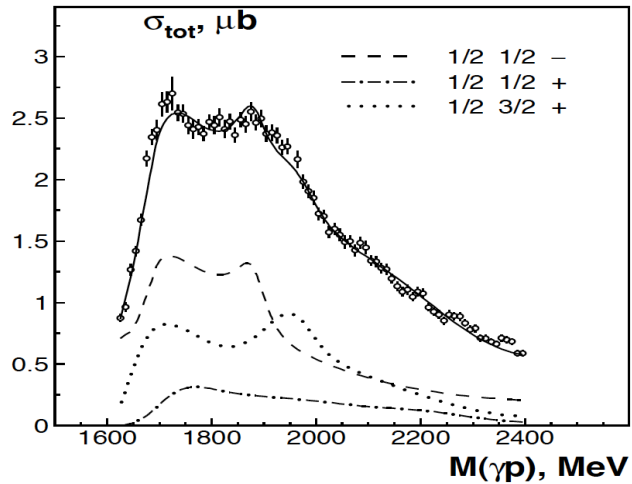}}
\caption{\protect Integrated cross sections of the CLAS $\gamma p \to K^+\Lambda$ data.  
The curves represent fits by the BnGa group\cite{BnGa-1} to the differential cross sections\cite{mccracken}. The fit parameter 
are then used to construct the total cross section (solid line). Resonance structures are clearly visible. 
The broken curves represent s- and p- wave contributions from the fit indicating the presence of $J^P$ states with 
$1/2^-$, $1/2^+$ and $3/2^+$. } 
\end{center}
\label{KLambda-tot}
\end{figure}
With all 4 complex production amplitudes known, a single channel energy-independent and model-independent 
partial wave analysis will be possible, e.g. using the techniques of Argand diagrams in the search for resonant states. 
The energy-dependence of a partial-wave amplitude is influenced by other reaction channels due 
to unitarity constraints. To fully describe the energy-dependence of an amplitude other channels have to be included in a 
coupled-channel analyses. Such an analysis has been developed by the Bonn-Gatchina (BnGa) group\cite{BnGa-1}, also
at JLab\cite{hlee2007}, and at Juelich\cite{krewald}. The BnGa group included nearly all available 
photo production data into their most recent analysis\cite{boga2012}.  
\vspace{-0.3cm}
\begin{figure}[t]
\resizebox{0.95\columnwidth}{!}{\includegraphics{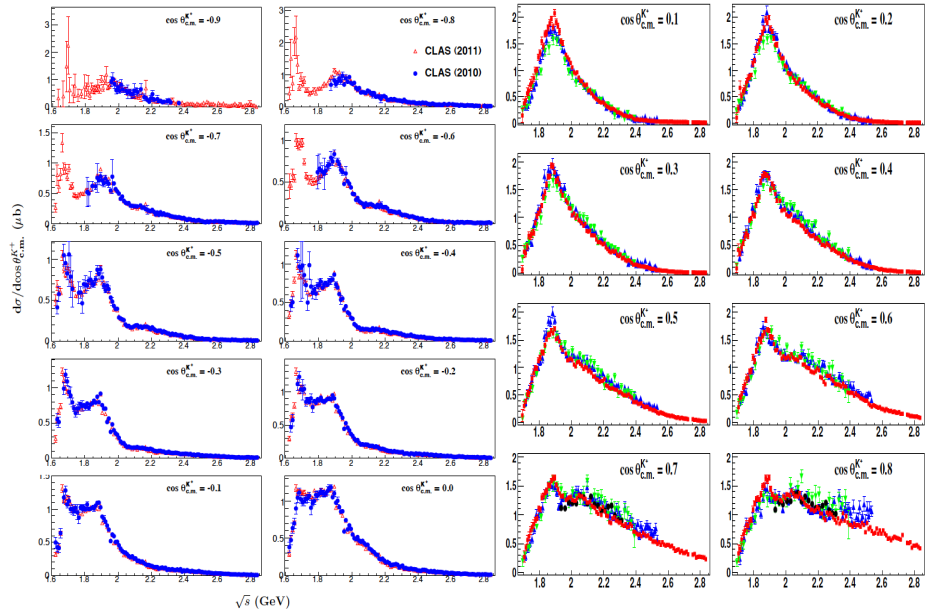}}
\vspace{-1.5cm}
\caption{Left panel: Invariant mass dependence of the $\gamma p \to K^+\Lambda$ differential cross section for the backward
hemisphere in 
cms polar angle $\cos(\theta)$. The blue full circles are based on the topology $K^+p\pi^-$, the red open triangles
are based on topology $K^+p$ or $K^+\pi^-$, which has extended coverage towards lower W at backward angles
and allows better access to the resonant structure near threshold. Right panel: Invariant mass dependence of the
 $\gamma p \to K^+\Sigma^\circ$ differential cross section for the forward hemisphere in 
cms polar angle $\cos(\theta)$. The blue triangles are the 2006 CLAS data, the red squares are the 2010 CLAS results 
with extended mass coverage, the green triangles are SAPHIR data, and the black circles are LEPS results covering
only forward angles. }
\label{KLambda-crs}
\end{figure}

\vspace{0.5cm}
The data sets with the highest impact on resonance amplitudes in the mass range from 1.8 to 2.2 GeV have been 
kaon-hyperon production using a spin-polarized photon beam and 
where the polarization of the $\Lambda$ or $\Sigma^\circ$ is also measured. The high precision cross section and 
polarization data from CLAS\cite{bradford1,mcnabb,mccracken,bdey2010,bradford2} provide nearly 
full polar angle coverage and span the $K^+\Lambda$ and $K^+\Sigma^\circ$ invariant mass range from threshold to 2.9 GeV, 
hence covering the full nucleon resonance domain where new states may be discovered. 
The integrated $\gamma p \to K^+\Lambda$ cross section is shown in Fig.1. The invariant mass $M(\gamma p)$ 
of the $K^+\Lambda$ channel shows two resonance-like structures near 1.7 GeV and near 1.9 GeV. The low mass structure is due to well-known p-wave states, $N(1710)1/2^+$ and $N(1720)3/2^+$, while the second structure has been subject to 
various interpretation as a $N(1900)3/2^-$,  $N(1900)1/2^+$, or $N(1900)3/2^+$, or a combination of several states. 
While the integrated cross section gives hints of possible new states, more details may be gleaned from the 
differential cross sections shown in Fig.\ref{KLambda-crs} for $K^+\Lambda$ and $K^+\Sigma^\circ$ production. 
\begin{figure}[t]
\hspace{-0.5cm}
\vspace{-0.5cm}
\resizebox{1.0\columnwidth}{!}{\includegraphics{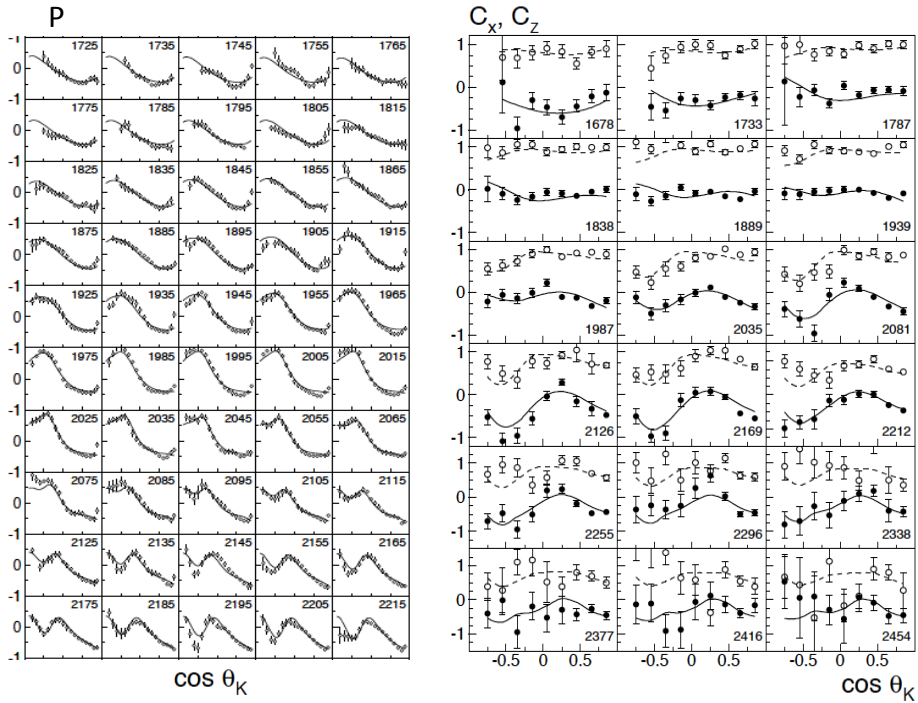}}
\caption{Polar angle dependence of the $\Lambda$ 
recoil polarization (left), and the polarization transfer from the circularly polarized photon to the recoil $\Lambda$ (right). 
The curves represent the Bonn-Gatchina analysis~\cite{boga2012}, using a coupled channel approach. 
The entire energy and polar angle ranges are fitted simultaneously.}    
\label{bogafit}
\end{figure}

The $K^+\Lambda$ differential cross section data show resonance-like structures at 1.7 GeV and 1.9 GeV that are particularly 
prominent and well-separated at backward angles, while at more forward angles t-channel processes are more prominent and 
dominate the cross section. Still, the characteristics of resonance production is present at all kinematics, even at the
very forward angles. The $K^+\Sigma$ channel also shows significant resonant behavior. The peak structure at 1.9 GeV is 
present at all angles with a maximum strength near 90 degrees, consistent with the behavior of a $J^P= 3/2^+$ 
p-wave. 
\begin{table}[h]
\begin{center}
\caption{Evidence for excited nucleon states from hyperon, $p\omega$ and $p\eta^\prime$ photoproduction. Asterix in 
bold face indicate new entries in PDG2012, or known states with upgraded rating or states that have been confirmed 
in the BnGa analysis\cite{boga2012} by including photoproduction channels. The other columns show the sensitivity of 
different channels to specific states. The masses are slightly different from the BnGa analysis or PDG values.}
\label{tab:1}  
\begin{tabular}{llllllll}
\hline\noalign{\smallskip}
N(mass)$J^P$ & PDG 2012 & $K\Lambda$ & $K\Sigma$ & $\gamma$ p&p$\omega$& p$\eta^{\prime}$    \\
\noalign{\smallskip}\hline\noalign{\smallskip}
$N(1710)1/2^+$&  {\bf\large ***} & *** & ** & *** &  & \\
$N(1880)1/2^+$ &  {\bf\large **}  & ** &  & ** &  & \\
$N(2100)1/2^+$ &* & & & & & 2130 \\
$N(1895)1/2^-$ & {\bf \large **}  & ** & * & ** &  &  1920\\
$N(1900)3/2^+$ &  {\bf \large***}  & *** & ** & *** &  &  \\
$N(2040)3/2^+$ & *  &  &  &  &  & 2050 \\
$N(1875)3/2^-$  &   {\bf\large ***}  & *** & ** & *** &  & \\
$N(2150)3/2^-$  &  {\bf\large **}  & ** &  & ** &  & \\
$N(2000)5/2^+$ &  {\bf\large **}  & ** & * & ** & 1950& \\
$N(2060)5/2^-$ &  {\bf\large **}  &  & ** & ** & & 2080\\
\noalign{\smallskip}\hline
\end{tabular}
\end{center}
\label{new-states}
\end{table}
\begin{figure}[]
\hspace{1.5cm}
\resizebox{0.7\columnwidth}{!}{\includegraphics{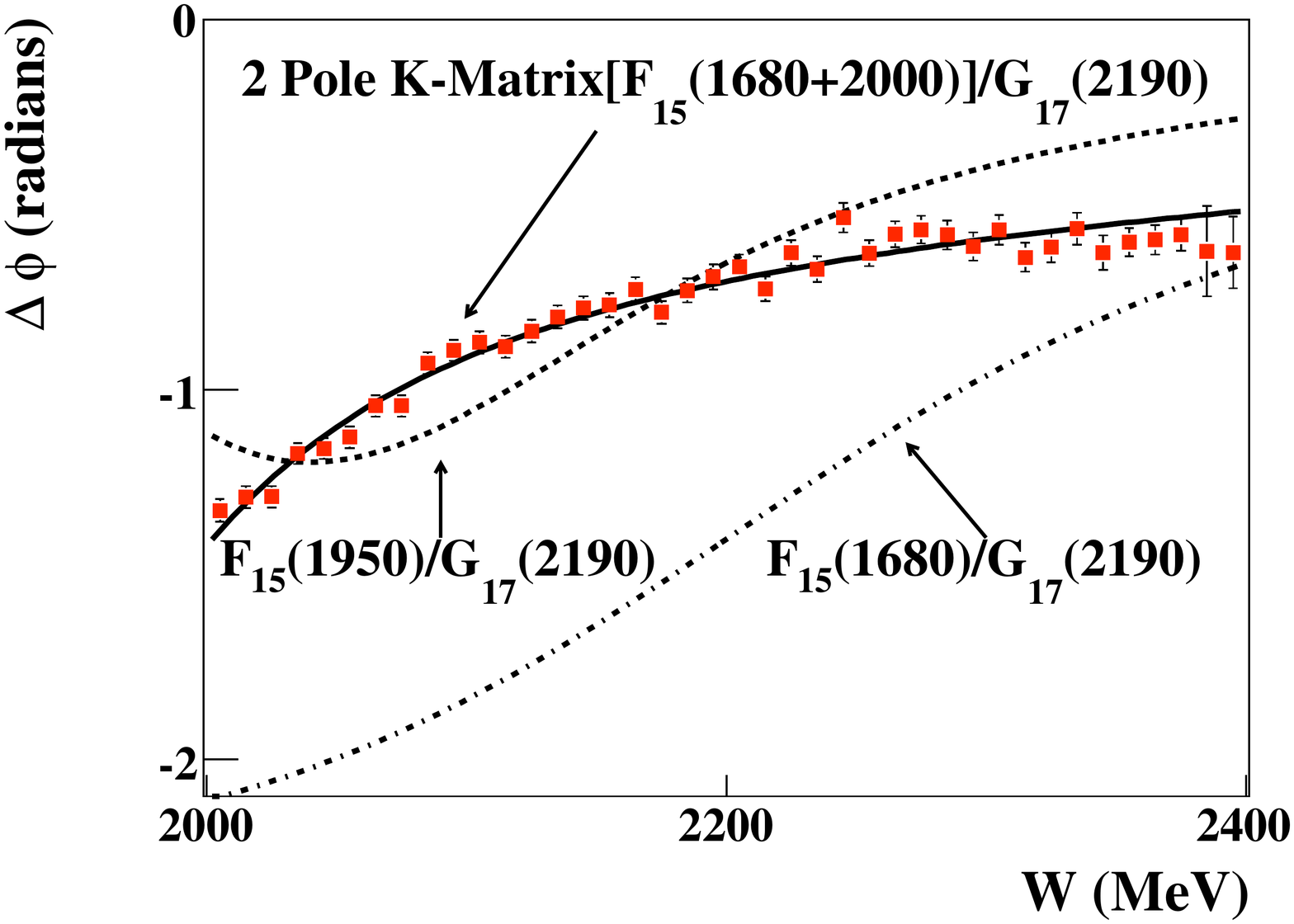}}
\caption{Fit results for $\Delta\phi=\phi_{7/2^-}-\phi_{5/2^+}$  vs 
the invariant $p\omega$ mass. The dot-dashed line is the phase motion expected 
 using constant width Breit-Wigner distributions and the parameters quoted by the PDG for the
  $N(1680)5/2^+$ and $N(2190)7/2^-$. The dashed line required the $J^P =7/2^-$ parameters to be 
within the PDG limits for the $N(2190)7/2^-$, while allowing the $J^P = 5/2^+$ parameters to vary freely. 
The solid line is obtained using a constant width Breit-Wigner distribution for the $N(2190)7/2^-$, 
and a 2-pole single channel K-matrix for the $J^P=5/2^+$ wave.}
\label{clasfit}
\end{figure}
The BnGa coupled channel analysis used these data together with hadronic 
scattering data and other photoproduction channels and identified five new candidate nucleon states, not previously seen with any significance, and confirmed two other states previously seen, but in need of confirmation.  Figure~\ref{bogafit} shows the BnGa fit to some of the $K^+\Lambda$ data on the 
$\Lambda$ recoil polarization, and the $\gamma \to \Lambda$ polarization transfer coefficient $C_x$ and $C_z$, which correspond to polarization transfer from circularly polarized photons to  
transverse and longitudinal $\Lambda$ polarization, respectively.  

The table~\ref{tab:1} shows the evidence for new 
isospin $I = {1\over 2}$ states and the ratings for newly observed states in the 2012 edition of the Review of Particle Properties\cite{pdg2012}. It should be noted that the photoproduction reactions have now much better precision than the older pion or kaon induced reactions. This is especially the case for the $K^+\Lambda$ and $K^+\Sigma^\circ$ final states measured at CLAS. There are also precision cross section data with extended coverage in photon energy on $p\pi^\circ$, $n \pi^+$, $p \eta$, $p\omega$, and for the first time high statistics data on $\gamma p \to p \eta^\prime$. It is noteworthy that the existence of one of
observed states, $N(1710)1/2^+$,  was recently challenged by the GWU PWA where the state was not observed 
in the analysis\cite{gwu} of $\pi N \to \pi N$ and $\gamma N \to \pi N$ reactions. 
The CLAS data for the $\gamma p \to K^+ \Lambda$ channel clearly show a strong peak near 1700 MeV which in the BnGa analysis 
represents states of $J^P$ with $1/2^+$ and $3/2^+$.  This illustrates that one has to go beyond pion production in the quest for new excited states. The table also includes evidence for some of the states from new single channel analyses. The CLAS collaboration has recently published high precision data on $\gamma p \to p \omega$ including a single channel energy-independent event-based analysis in the mass range from threshold to 2400 MeV\cite{mwilliams2}. This type of analysis is a first for baryons. The energy dependence of the partial wave is found from the data and not a-priory imposed on the fit. The phase motion shown in Fig.\ref{clasfit} requires existence of a $N(1950)5/2^+$ state. 

The precision CLAS $\gamma p \to p\eta^\prime$ data\cite{mwilliams3} have been analyzed using a relativistic quark model approach\cite{qiang}, and independently in an effective Lagrangian approach\cite{haberzettl}. Both analyses give evidence for the $N(1895)1/2^-$ state also seen in the BnGa analysis  of the $K\Lambda$ and $K\Sigma$ data. The analysis of ref.\cite{qiang} shows evidence for the $N(2060)5/2^-$ seen also in the BnGa analysis albeit at a slightly higher mass, while analysis\cite{haberzettl}, which is limited to s- and p-waves, shows evidence for a $J^P=1/2^+$ state at 2130 MeV, which could be identified as the poorly established $N(2100)1/2^+$. An additional $3/2^+$ state at a mass of 2050 MeV could be assigned to the one star  $N(2040)3/2^+$.    

\section{Conclusions and Outlook}
The emergence of significant new evidence for several excited states of the nucleon with masses in the range 1900 - 2150~MeV is largely 
due to the precise new photoproduction data for channels such as $K^+\Lambda$ and $K^+\Sigma^\circ$, and to some degree to new $p\eta$, $p\omega$ and $p\eta^\prime$ data. Most of the states have been identified in the BnGa multi-channel analyses, and several have been seen in single channel analyses. This marks indeed a milestone and demonstrates the importance of the electromagnetic interaction as a probe, with the added sensitivity from the polarization observables that have been measured. Eventually one would like to see all resonances identified in fully coupled-channel analyses, where the 
partial wave contributions are extracted in an energy-independent fashion. Until this becomes routine, single channel analyses have an important role to play. It is much easier to control systematics for single reactions then for a multitude of channels coming from different experiments with different systematics. Also, it is much easier to have "complete" experimental information for a single channel that allows for a model-independent extraction of the scattering amplitudes. Ideally, one would like to see all new states, as well as some of the "established" states, discovered (or rediscovered) in single channels, e.g. using the Argand diagram technique, as well as in coupled channel partial wave analyses. With the new precise polarization data from the various photoproduction experiments this may become a reality in the not too distant future.  

\vspace{0.3cm}
Authored by Jefferson Science Associates, LLC under U.S. DOE Contract No. DE-AC05-06OR23177. 
The U.S. Government retains a non-exclusive, paid-up, irrevocable, world-wide license to publish or 
reproduce this manuscript for U.S. Government purposes.

\end{document}